\documentclass[twocolumn,prl,showpacs]{revtex4}
\usepackage{epsf,graphicx,amssymb}

\usepackage{epstopdf}
\usepackage{amsmath}  
\usepackage{amssymb}
\usepackage{bbold}
\usepackage{mathbbol}
\usepackage{amscd}
\usepackage{amsfonts} 
\usepackage{graphicx} 
\usepackage{color}

\usepackage{amssymb,amsmath,amsthm}
\usepackage[colorlinks=true,linkcolor=blue,citecolor=blue]{hyperref}

\begin{document}
\title{Hysteretic Faraday waves}
\author{Nicolas P\'erinet}
\email{nperinet@ing.uchile.cl}
\author{Claudio Falc\'on}
\address{Departamento de F\'isica, Facultad de Ciencias F\'isicas y Matem\'aticas, Universidad de Chile, Casilla 487-3, Santiago, Chile}
\author{Jalel Chergui}
\author{Damir Juric}
\address{LIMSI, CNRS, Universit\'e Paris-Saclay,  B\^at 508, rue John von Neumann, Campus Universitaire, F-91405 Orsay }
\author{Seungwon Shin}
\address{Department of Mechanical and System Design Engineering, Hongik University, Seoul, 121-791 Korea}
\date{\today}

\begin{abstract}

We report on the numerical and theoretical study of the subcritical bifurcation of parametrically amplified waves appearing at the interface between two immiscible incompressible fluids when the layer of the lower fluid is very shallow. As a critical control parameter is surpassed, small amplitude surface waves bifurcate towards highly nonlinear ones, with twice their amplitude. We propose a simple phenomenological model which can describe the observed bifurcation. We relate this hysteresis with the change of shear stress using a simple stress balance, in agreement with numerical results. 
\end{abstract}

\date{\today}
\maketitle

The Faraday experiment~\cite{Faraday1831} is a paradigmatic example in pattern forming systems, due to its simplicity and richness. Standing waves appear at the interface between two fluids as they are vibrated vertically with a certain amplitude larger than a threshold at a given frequency. Extended Faraday waves have been observed with various types of fluids~\cite{WagnerPRL1998,RaynalEPJB1999,AransonRMP1006,BallestraJNonFMech2007} on a wide range of physical configurations~\cite{FauveJPhys1991,TrinhPRL1996,KudrolliPRE2001,FalconEPL2007,PucciPRL2011}, displaying numerous shapes~\cite{ChristiansenPRL1992,FauvePRE1993,KudrolliPhysD1998,SilberPRL2012}. Localized structures have also been observed in the Faraday experiment in several situations~\cite{RudnickPRL1984,MeloNature1996, FinebergPRL1999,FinebergPRL2000,MerktPRL2004,RachjebachPRL2011,FalconELP2012}. The description of the nature, origin and dynamical properties of Faraday waves has been the focus of a large scientific endeavor since the seminal work of Faraday. A particular property of these waves is their ability to display coexistence (bistability) between different wave states. Several attempts have been made to describe this feature, either from first principles~\cite{ChenPRL1997,ZhangJFM1997,SkeldonSIAM2007,SkeldonJFM2015} or phenomenological standpoints (see~\cite{CrossHohenbergRPM1993} and references therein), although with little or no real connection to a dynamical or structural change of the wave pattern. 

In this Letter, we report on the numerical simulation of hysteretic Faraday waves at the interface between two immiscible and incompressible fluids, where the lower fluid layer is very shallow. Two branches of Faraday waves are observed with different amplitudes and shapes characterized by the surface deformation and velocity field of both fluids. We propose that this hysteretic jump is related to a sudden shift in the localization of the viscous boundary layer which moves from the interface to the bottom of the cell. We explain physically the observed hysteresis by a balance of the stresses exerted to the lower fluid layer.

We simulate numerically the equations governing the motion of two incompressible and immiscible fluids, separated by a sharp interface, using a single fluid formulation  
\begin{equation}
\label{eq-NS}
\rho\frac{D\textbf{u}}{Dt}=-\nabla p+\rho \textbf{G}+\nabla\cdot\mu\left(\nabla\textbf{u}+\nabla\textbf{u}^{\rm T}\right)+\textbf{F},
\end{equation}
where $\textbf{u}$ is the velocity field satisfying $\nabla\cdot\textbf{u}=0$, $D/Dt=\left(\partial_t +\textbf{u}\cdot\nabla\right)$ is the material derivative and $p$ is the pressure. The density $\rho$ and dynamic viscosity $\mu$ remain constant within each phase. $\textbf{F}$ and $\rho\textbf{G}$  stand for the densities of surface tension forces located at the interface and of volume forces, respectively. Here  $(\cdot)^T$ denotes the transposition operator. In the frame of reference of the vibrating fluids, $\textbf{G}=-(g+a\cos(\omega t))\textbf{e}_z$, where $g$ is gravity, $a$ the gravity modulation amplitude and $\omega=2\pi/T=2\pi f$ its angular frequency. The vector $\textbf{e}_z$ is oriented vertically and points upwards. The force density $\textbf{F} = \sigma \mathcal{K}\nabla I$ depends on the surface tension coefficient $\sigma$ which remains constant over the interface, the interface mean curvature $\mathcal{K}$ and an indicator function $I(x,y,z,t)$ that takes the value 0 in the heavier phase and 1 otherwise. 
\begin{figure}
\centering
\includegraphics[width=\columnwidth, height=1.22\columnwidth]{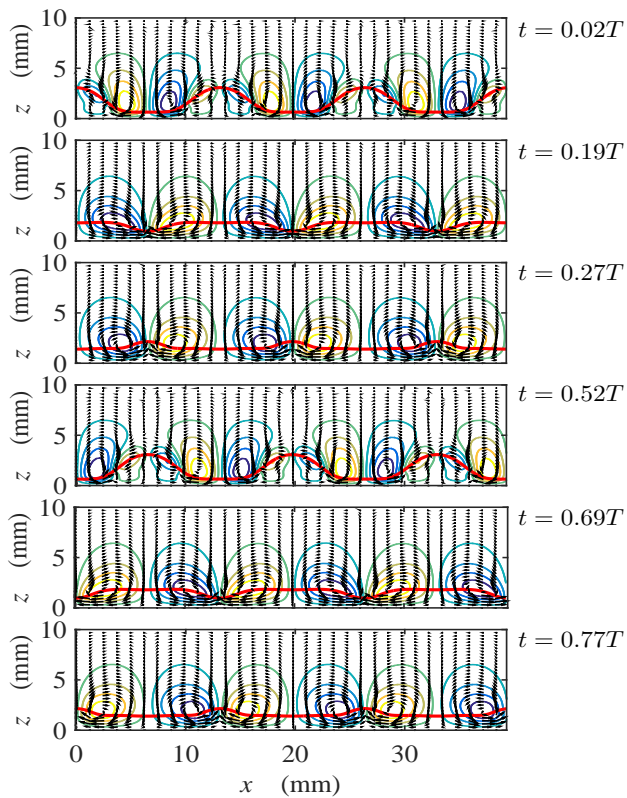} 
\caption{(color online) Temporal snapshots of the interface deformation $\zeta(x,t)$ (thick line), velocity field $\textbf{u} (x,z,t)$ (arrows) and stream function (contour lines) at $a=$ 39.5 m/s$^{2}$ (lower branch).}
\label{FigLower}
\end{figure}

\begin{figure}
\centering
\includegraphics[width=\columnwidth, height=1.22\columnwidth]{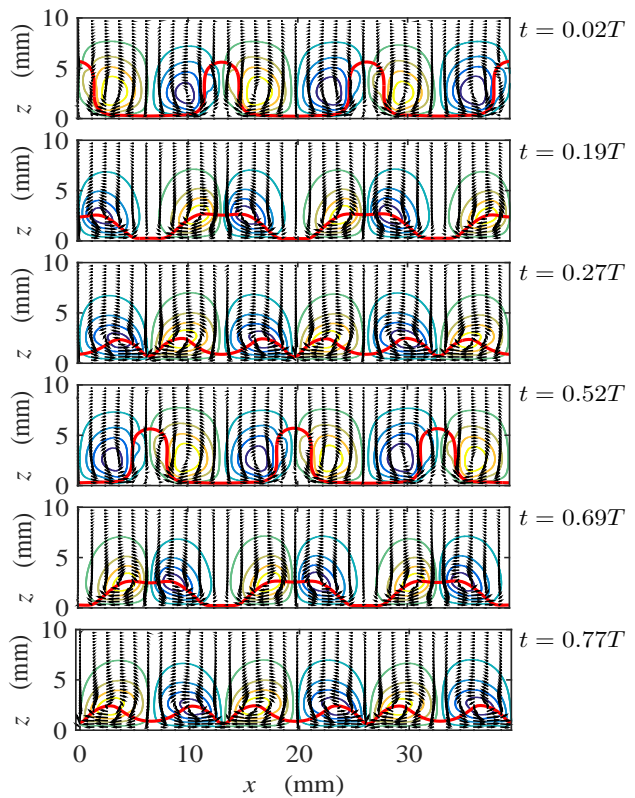} 
\caption{(color online) Same snapshots as Fig.~\ref{FigLower} at $a=$ 39.5 m/s$^{2}$ (upper branch).}
\label{FigUpper}
\end{figure}

The problem is treated using a massively parallel numerical code explained in Refs.~\cite{NicoPJFM2009,Shin-jcp-2007,Shin-ijnmf-2009,Shin-cf-2014} that can simulate Faraday waves in big domains \cite{NicoPJFM2015}. The simulated domain has a height $H=$ 10 mm. The thickness of the lower fluid layer is $h=$ 1.6 mm. The physical parameters are taken from Ref.~\cite{kemw-pre-2005}. The heavy fluid has density $\rho_1=1346$ kg/m$^3$ and dynamic viscosity $\mu_1= 7.2\times10^{-3}$ Pa s while for the lighter fluid $\rho_2=949$ kg/m$^3$ and $\mu_2=2.0\times10^{-2}$ Pa s. The surface tension is $\sigma=35$ mN/m. The modulation frequency $f$ is 12 Hz and its amplitude $a$ is varied. $\textbf{u}$ is subjected to no-slip boundary conditions at the top and bottom walls of the domain and is horizontally periodic.

The linear analysis~\cite{Kumar-jfm-1994} shows that using the above parameters the critical wavelength is $\lambda_c=13.2$ mm and the critical amplitude is $a_c=25.8$ m/s$^2$. The dimensions of the box in the system of coordinates $(x,y,z)$ are $L\times W\times H=39.6\times3.30\times10$ mm$^3$. The domain contains exactly three critical wavelengths longitudinally in order to spot eventual large scale effects at high $a$. The $y$ transverse dimension of the domain is small enough for the flow to remain essentially two-dimensional in the $x-z$ plane, which we have checked numerically.  The numerical resolution used in our simulation runs is $128\times8\times128$. We have checked that the same phenomena are observed with higher resolutions. Numerical stability and accuracy are assured using a dynamically bounded time step $\Delta t$ ~\cite{NicoPJFM2015}. To simplify our analysis, we restrict ourselves to the two-dimensional dynamics of surface waves described by the interface deformation $\zeta(x,t)$ and velocity field $\textbf{u}(x,z,t)$ (as shown in Fig.~\ref{FigLower} and~\ref{FigUpper}).

\begin{figure}[t]
\centering
\includegraphics[width=1.0\columnwidth]{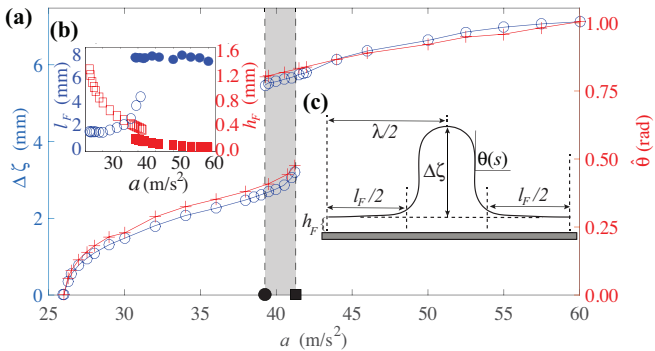} 
\caption{(color online) (a) Bifurcation diagram for $\Delta\zeta$ ($\circ$) and $\hat{\theta}$ ($+$) as a function of $a$ for the weakly (lower branch) and highly (upper branch) nonlinear saturated wave. Hysteresis occurs in the shaded region: $a_c^{d}=$ 39.25 m/s$^{2}$ ($\bullet$) and $a_c^{u}=$ 41.25 m/s$^{2}$ ($\blacksquare$) are displayed in the acceleration axis. (b) $l_F$ ($\circ$) and $h_F$ ($\square$) as a function of $a$ on the lower (open symbols) and upper (full symbols) branches of the bifurcation diagram. (c) Surface deformation $\zeta(x,t)$ (continuous line) at $a$ = 39.5 m/s$^{2}$ showing the definition of $\theta(s)$ and $\Delta\zeta$ for the upper branch of the bifurcation diagram.}
\label{FigAmplitud}
\end{figure}

When $a$ surpasses the critical value $a_c^{p}=26$ m/s$^{2}$, the flat surface becomes unstable to infinitesimally small perturbations and stationary subharmonic surface gravity waves appear with a wavelength $\lambda_c=L/3$. Following Ref.~\cite{NicoPJFM2009}, we show the bifurcation diagram of the saturated surface wave peak-to-peak amplitude $\Delta\zeta$ in Fig.~\ref{FigAmplitud}(a). $\Delta\zeta$ shows the same distinctive features as the Fourier mode amplitude at $\lambda_c$ and is straightforward to measure. Its dependence on $a$ can be accurately fitted using the reduced control parameter $\epsilon=(a-a_c^{p})/a_c^{p}$ as $\Delta\zeta\sim\epsilon^{1/2}$ for $\epsilon<0.6$. As we increase $a$ further than 40 m/s$^{2}$ the dependence of $\Delta\zeta$ on $\epsilon$ changes and a slight curvature towards larger values of $\Delta\zeta$ appears on the bifurcation diagram. As $a$ exceeds $a_c^u$=41.25 m/s$^{2}$ a secondary instability occurs: $\Delta\zeta$ increases by a factor 2 and the shape of $\zeta(x,t)$ becomes highly nonlinear, displaying localized peaks and almost horizontal troughs of constant length $l_F\simeq$ 8 mm. The thickness of these troughs at $a_c^{u}$ in the upper branch of the bifurcation, $h_F\simeq$ 0.25 mm, is twice as small as the one on the lower branch, as depicted in Fig.~\ref{FigAmplitud}(b). These variables are defined geometrically in Fig.~\ref{FigAmplitud}(c). As a consequence of mass conservation while $\Delta\zeta$ becomes larger $h_F$ becomes smaller. In the upper branch $\zeta(x,t)$ also becomes multivalued. To avoid misinterpretations of the bifurcation diagram by following $\Delta\zeta$ as $\zeta(x,t)$ becomes multivalued in the upper branch, we have also used the angle $\theta(s)$ that the local tangent to the interface makes with the $x$ axis at normalized arc length $s$ as an order parameter (see Fig.~\ref{FigAmplitud}(c)). $\theta(s)$ is single valued for all values of $a$. We have computed the Fourier amplitude at $\lambda_c$ of $\theta(s)$ in arc length $s$, $\hat{\theta}$, which displays the same bifurcation diagram as $\Delta\zeta$ (see Fig.~\ref{FigAmplitud}(a)). A hysteresis loop is displayed as this state is sustained decreasing $a$ until $a_c^{d}$ = 39.25 m/s$^{2}$$\neq a_c^{u}$ (see Fig.~\ref{FigAmplitud}(a)). To wit, we show in both Fig.~\ref{FigLower}(a) and Fig.~\ref{FigLower}(b) $\zeta(x,t)$ and $\textbf{u}(x,z,t)$ for the same value of $a=$ 39.5 m/s$^{2}$.

\begin{figure}[b]
\centering
\includegraphics[width=\columnwidth]{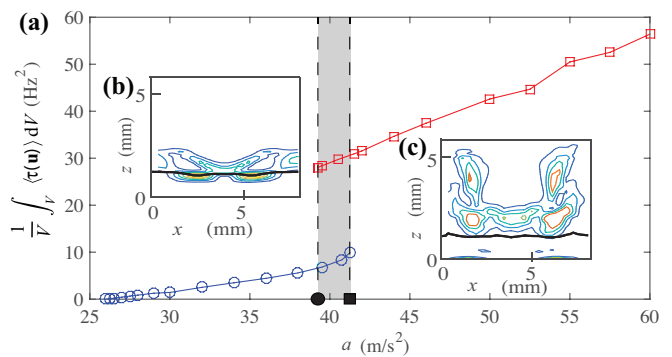} 
\caption{(color online) (a) Bifurcation diagram for the space-averaged dissipation rate of the mean velocity $\frac{1}{V}\int_{V}\langle\tau(\textbf{u})\rangle$ as a function of $a$ for the weakly ($\circ$) and highly ($\square$) nonlinear saturated wave. Hysteresis occurs in the shaded region: $a_c^{d}=$ 39.25 m/s$^{2}$ ($\bullet$) and $a_c^{u}=$ 41.25 m/s$^{2}$ ($\blacksquare$) are displayed in the acceleration axis. Insets: Contour lines of $\tau(\langle \textbf{u}\rangle)$ for $a=$39.5 m/s$^{2}$ in the lower (b) and upper (c) branches. The brighter the lines, the higher the dissipation. The thick dark line is the interface. }
\label{FigTau}
\end{figure}

Hysteresis is also reflected in the changes of the velocity field properties, specifically, how the energy dissipation rate $\tau(\textbf{u})=\mu\sum_{i,j=1}^{2}(\partial \textbf{u}_i/\partial x_j)^{2}$ evolves as $a$ increases. We first focus on the mean dissipation rate $\langle\tau(\textbf{u})\rangle$. Here $\langle\cdot\rangle$ stands for time average. The amplitude jump and hysteresis loop observed for $\Delta\zeta$ and $\hat{\theta}$ are also observed for the space-averaged mean dissipation rate $\frac{1}{V}\int_{V}\langle\tau(\textbf{u})\rangle dV$ as shown in Fig.~\ref{FigTau}(a). A roughly linear dependence on $a$ can be observed, with different slopes for lower and upper branches. A spatial change in the structure of $\tau$ can be also observed by measuring the dissipation rate of the mean velocity $\tau(\langle \textbf{u}\rangle)$.  For the lower branch, the largest values of $\tau(\langle \textbf{u}\rangle)$ are localized at the interface where the shearing of both fluids is the strongest. The profile of $\tau(\langle \textbf{u}\rangle)$ changes in the upper branch: $\tau(\langle \textbf{u}\rangle)$ presents maxima at the interface and also at the bottom of the domain. When $a$ is further increased, $\tau(\langle \textbf{u}\rangle)$ becomes localized at the bottom of the cell where the viscous boundary layer dissipates the largest part of the kinetic energy of the flow. The difference in structure of $\tau(\langle \textbf{u}\rangle)$ for $a$ = 39.5 m/s$^{2}$ is shown in Fig.~\ref{FigTau}(b)-(c). All temporal Fourier components of $\tau$ display the same bifurcation diagram as $\langle\tau(\textbf{u})\rangle$ (not plotted here), showing that the structural change is a global one. 

To explain this hysteretic transition and the qualitative changes reported above at the transition we present a physical explanation relating the above data. We propose that the amplitude jump and wave hysteresis can be understood from a balance between lubrication and hydrostatic stresses \cite{FalconELP2012,Reynolds} that is coupled with a change of the flow regime within the film.

From the data displayed in Fig.~\ref{FigAmplitud}(a), the depth of the layer reduces to $h_F=h-\Delta\zeta/2\simeq0.5$ mm at $a_c^{u}$ on the lower branch of the bifurcation, occurring roughly over $\lambda_c/2$ (see Fig.~\ref{FigLower} and Fig.~\ref{FigAmplitud}(c)). At this point the Reynolds number $Re^{u}=\rho_1\omega h_F^{2}/\mu_1\simeq4$ and a transition occurs as the viscous stress within the film increases, forcing the hydrostatic stress to do the same. Then, a new equilibrium state is reached in the upper branch, with a thinner $h_F$ and a larger $l_F$ (see Fig.~\ref{FigUpper} and Fig.~\ref{FigAmplitud}(b)). In this upper branch, the flow in the thin film is described by Stokes dynamics $\nabla p=\mu\nabla^2\textbf{u}$, assuming that $\partial_z\gg\partial_x$, $u\gg w$ and $p$ is constant along the $z$ direction, where $u$, $w$ and $p$ stand for the horizontal and vertical components of the velocity $\textbf{u}$ and the pressure, respectively. Hence
\begin{equation}
\partial_{zz}u=\frac{\Delta p}{\mu_1 l_F},\qquad\nabla^2 w=0,
\label{eq-Stokes}
\end{equation} 
where $\Delta p$ is the difference between the pressure in the lower fluid under the column and inside the film. Assuming zero velocity at the bottom of the domain $z=0$ and a stress-free interface $\partial_z u=0$ at $z=h_F$, the solution of (\ref{eq-Stokes}) is 
\begin{equation}
u=\frac{\Delta p}{\mu_1 l_F}\left(\frac{z^2}{2}-h_F z\right),
\end{equation}
which is averaged over the fluid height to evaluate the viscous stress $\sigma_s=3\mu_1 l_F \bar{u}/h_F^2$. This stress
makes the film resist detaching when a viscous regime is achieved at $Re=\rho_1\omega h_F^{2}/\mu_1<Re^{d}\simeq1$. The critical Reynolds number $Re^{d}$ is calculated for $h_F$ at the transition from the upper to the lower branch in Fig.~\ref{FigAmplitud}(a) and it sets a critical depth $h_0\sim0.25$ mm. From dimensional analysis, one expects $ \bar{u}\sim \omega l_F$, which is confirmed by our numerical simulations as $ \bar{u}/\omega l_F\simeq1/12$ in the upper branch. Hence, $\sigma_s\simeq\mu_1 l_F^2 \omega/(4h_F^2)$ and the force per unit length arising from this shear stress, $F_s\simeq\mu_1\omega l_F^{3}/(12h_F^{2})$, must compensate the hydrostatic pressure contribution $F_h$ from both ends of the film at maximum acceleration, to ensure the film sustainment. The stress balance reads $F_s=2F_h$ where $F_h$ is given at each border of the film by the difference of hydrostatic pressure between the zone inside the film and the zone outside the film: $F_h\simeq(\rho_1-\rho_2)(a^{*}+g)\Delta\zeta^{2}$, $a^{*}$ denoting the maximum acceleration. After the transition, $F_s$ and $F_h$ are estimated at 300 mN/m in the upper branch, which is 10 times larger than the stress contribution from surface tension that is consequently neglected. Using mass conservation $h\lambda_c\simeq\Delta\zeta(\lambda_c-l_F)$ to relate $\Delta\zeta, h_F$ and $l_F$, the balance reads
\begin{equation}
l_F^3(\lambda_c-l_F)^2 \simeq \frac{6 g h^2\lambda_c^2 Re}{\omega^2}\left(1+\frac{a^*}{g}\right)\left(1-\frac{\rho_2}{\rho_1}\right),
\end{equation}
where $Re<Re^{d}$, according to our simulation results on the upper branch (lubrication flow) and $\omega$ is given by the dispersion relation for shallow waves
\begin{equation}
\omega^2=\left(\frac{\rho_1-\rho_2}{\rho_1+\rho_2}\right)\frac{16\pi^2}{\lambda_c^2}gh,
\end{equation}
with the forcing frequency $\omega$ is twice that of the fluid's response.
At $Re=Re^d$, the stress balance takes the form
\begin{equation}
l_F^{3}(\lambda_c-l_{F})^{2}\simeq\frac{3h\lambda_c^{4}}{8\pi^{2}}\left(1+\frac{a^{*}}{g}\right)\left(1+\frac{\rho_2}{\rho_1}\right).
\end{equation}
The left-hand side of this balance reaches its maximum when $l_F=3\lambda_c/5\simeq7.9$ mm, close to the constant value found in our simulations for $l_F$ in the upper branch. Thus $\Delta\zeta\simeq5$ mm and the stress balance is achieved when $a^{*}/g\simeq3.4$. As $l_F$ rests constant in the upper branch, when $a>a^{*}$, $h_F$ decreases to compensate the increase in $F_h$. When $a<a^{*}$, $h_F>h_0$ ($Re>1$), changing the nature of the flow from lubrication ($\sigma_s\sim l_F^{2}/h_F^{2}$) to viscous flow ($\sigma_s\sim l_F/h_F$). With this functional change $F_s$ cannot sustain $F_h$, as it is $l_F/h_F\sim30$ times smaller. We conjecture that this shift in magnitude for $\sigma_s$ as the film changes its thickness is the reason for the observed amplitude jump.

In summary, using numerical simulations, we have observed the hysteretic bifurcation of Faraday waves in very shallow layers. The coexistence between weakly and highly nonlinear Faraday waves, observed experimentally, is now confirmed numerically and understood as the bifurcation of a wave pattern presenting hysteresis, which we conjecture is related to the structural shift of the viscous shear rate. The loop mechanism can be understood by stress balance between hydrostatic and shear stresses, which changes as the depth of the layer becomes shallower than the boundary layer of the fluid. This simple treatment is a first approach towards a more profound understanding of patten selection where dynamic changes in dissipation or structure are present, and opens questions related to the nature and origin of the hysteresis loop in shallow layers of Newtonian and complex fluids. 

We would like to thank L. Tuckerman and M. G. Clerc for fruitful discussions and to acknowledge the financial support of FONDECYT grants 1130354 and 3140522 as well as the Basic Science Research Program through the National Research Foundation of Korea (NRF) funded by the Ministry of Science, ICT and future planning (NRF-2014R1A2A1A11051346). This research was supported by the supercomputing infrastructure of the NLHPC (ECM-02) and GENCI (IDRIS).


\begin{thebibliography}{29}

\bibitem{Faraday1831} M. Faraday, Proc. Soc. London {\bf 121} 299Ð318 (1831).

\bibitem{WagnerPRL1998} C. Wagner, H. W. M\"uller and K. Knorr, Phys. Rev. Lett. {\bf 83}  308 (1998).

\bibitem{RaynalEPJB1999} F. Raynal, S. Kumar and S. Fauve, Eur. Phys. J. B {\bf 9} 175 (1999)

\bibitem{AransonRMP1006} I. Aranson and L. Tsimring, Rev. Mod. Phys. {\bf78}  641 (2006).

\bibitem{BallestraJNonFMech2007} P. Ballesta and S. Manneville, J. Non-Newtonian Fluid Mech. {\bf 147}  23 (2007)

\bibitem{FauveJPhys1991} S. Fauve, S. Douady and O. Thual, J. Phys. II {\bf1}(3), 311-322 (1991).

\bibitem{TrinhPRL1996} R. Glynn Holt and E. H. Trinh, Phys. Rev. Lett. {\bf77} 1274 (1996).

\bibitem{KudrolliPRE2001} A. Kudrolli, M. C. Abraham and J.P. Gollub, Phys. Rev. E {\bf63}, 026208 (2001).

\bibitem{FalconEPL2007} C. Falc\'on, U. Bortolozzo, E. Falcon and S. Fauve, EPL {\bf86} 14002 (2009).

\bibitem{PucciPRL2011} G. Pucci, E. Fort, M. Ben Amar and Y. Couder, Phys. Rev. Lett. {\bf 106}, 024503 (2011).

\bibitem{ChristiansenPRL1992} B. Christiansen, P. Alstr\o m and M. T. Levinsen, Phys. Rev. Lett. {\bf 68}, 2157 (1992).

\bibitem{FauvePRE1993} W. S. Edwards and S. Fauve, Phys. Rev. E {\bf 47}, R788 (1993).

\bibitem{KudrolliPhysD1998} A. Kudrolli, B. Pier and J. P. Gollub, Physica (Amsterdam) {\bf123D}, 99 (1998).

\bibitem{SilberPRL2012} A. M. Rucklidge, M. Silber and A. C. Skeldon, Phys. Rev. Lett. {\bf108}, 074 504 (2012).

\bibitem{RudnickPRL1984} J. Wu, R. Keolian and I. Rudnick
Phys. Rev. Lett. {\bf52}, 1421 (1984).

\bibitem{MeloNature1996} P. B. Umbanhowar, F. Melo and H. L. Swinney,  Nature {\bf382}  793Ð796 (1996).

\bibitem{MerktPRL2004}  F. S. Merkt, R. D. Deegan, D. I. Goldman, E. C. Rericha and H. L. Swinney, Phys. Rev. Lett {\bf92}, 184501  (2004).

\bibitem{FinebergPRL1999}O. Lioubashevski, Y. Hamiel, A. Agnon, Z. Reches,  and J. Fineberg, Phys. Rev. Lett. {\bf83} 3190 (1999).

\bibitem{FinebergPRL2000} H. Arbell and J. Fineberg, Phys. Rev. Lett. {\bf85} 756 (2000).

\bibitem{RachjebachPRL2011} J. Rajchenbach, A. Leroux and D. Clamond
Phys. Rev. Lett. {\bf107}, 024502 (2011).

\bibitem{FalconELP2012} C. Falc\'on, J. Bruggeman, M. Pasquali and R. D. Deegan, EPL {\bf98}, 30006 (2012).

\bibitem{ChenPRL1997} P. Chen and J. Vi\~nals, Phys. Rev. Lett. {\bf 79} 2670 (1997). 

\bibitem{ZhangJFM1997} W. Zhang and J. Vi\~nals, J. Fluid Mech. {\bf336}, 301Ð330 (1997).

\bibitem{SkeldonSIAM2007} A. C. Skeldon and G. Guidoboni,  SIAM J. Appl. Math. {\bf67} 1064-1100 (2007).

\bibitem{SkeldonJFM2015} A. C. Skeldon and A. M. Rucklidge, J. Fluid Mech. {\bf 777} 604-632 (2015).

\bibitem{CrossHohenbergRPM1993} M. C. Cross and P. C. Hohenberg, Rev. Mod. Phys. {\bf65}, 851 (1993).

\bibitem{NicoPJFM2009} N. P\'erinet, D. Juric and L. S. Tuckerman, J. Fluid Mech. {\bf 635} 1-26 (2009).


\bibitem[Shin(2007)]{Shin-jcp-2007}
S. Shin, J.~Comput.~Phys.~\textbf{222}, 872--878 (2007).

\bibitem[Shin \& Juric(2009)]{Shin-ijnmf-2009}
S. Shin and D. Juric, Int.~J.~Num.~Meth.~Fluids~\textbf{60}, 753--778 (2009).

\bibitem[Shin, Chergui \& Juric(2014)]{Shin-cf-2014}
S. Shin, J. Chergui and D. Juric, Comput.~Fluids, (submitted) arXiv:1410.8568[physics.flu-dyn].

\bibitem{NicoPJFM2015} L. Kahouadji, N. P\'erinet, L. S. Tuckerman, S. Shin, J. Chergui and D. Juric, J. Fluid Mech. {\bf772}, R2 (2015).







\bibitem{kemw-pre-2005} A. V. Kityk, J. Embs, V. V. Mekhonoshin and C. Wagner, Phys. Rev. E {\bf 72}, 036 209 (2005); {\bf79}, 029 902(E) (2009).

\bibitem[Kumar \& Tuckerman(1994)]{Kumar-jfm-1994}
K. Kumar and L.~S. Tuckerman, J.~Fluid Mech.~\textbf{279}, 49--68 (1994).









\bibitem{Reynolds} H. Ockendon H. and J. R. Ockendon, {\it Viscous Flow} (Cambridge University Press, New York, 1995).

\end{thebibliography}
\end{document}